\documentclass[superscriptaddress,twocolumn,showpacs,amsmath,amssymb,floats,pre]{revtex4-1}

\usepackage{graphicx}

\usepackage{dcolumn}

\usepackage{bm}

\usepackage{revsymb}

\usepackage[usenames]{color}

\usepackage{subfigure}

\begin{document}

\newcommand{\brm}[1]{\bm{{\rm #1}}}
\newcommand{\tens}[1]{\underline{\underline{#1}}}
\newcommand{\mm}{\overset{\leftrightarrow}{m}}
\newcommand{\xv}{\bm{{\rm x}}}
\newcommand{\kv}{\bm{{\rm k}}}
\newcommand{\Rv}{\bm{{\rm R}}}
\newcommand{\uv}{\bm{{\rm u}}}
\newcommand{\vv}{\bm{{\rm v}}}
\newcommand{\nv}{\bm{{\rm n}}}
\newcommand{\Nv}{\bm{{\rm N}}}
\newcommand{\ev}{\bm{{\rm e}}}
\newcommand{\dv}{\bm{{\rm d}}}
\newcommand{\bv}{\bm{{\rm b}}}
\newcommand{\lv}{{\bm{l}}}
\newcommand{\rv}{\bm{{\rm r}}}
\newcommand{\id}{\tens{\mathbb I}}
\newcommand{\bhv}{\hat{\bv}}
\newcommand{\bh}{\hat{b}}
\def\ten#1{\underline{\underline{{#1}}}}
\newcommand{\Ft}{{\tilde F}}
\newcommand{\Ftv}{\tilde{\mathbf{F}}}
\newcommand{\sigmat}{{\tilde \sigma}}
\newcommand{\sigmab}{{\overline \sigma}}
\newcommand{\ellv}{\mathbf{\ell}}
\newcommand{\qv}{\bm{{\rm q}}}
\newcommand{\pv}{\bm{{\rm p}}}
\newcommand{\tD}{\underline{D}}
\newcommand{\Tchange}[1]{{\color{red}{#1}}}
\newcommand{\Fa}{{\cal F}}
\newcommand{\Uv}{{\roarrow U}}
\newcommand{\Bv}{{\roarrow B}}
\newcommand{\Dt}{{\tensor {\cal D}}}
\newcommand{\cv}{{\mathbf c}}
\newcommand{\pt}{\tilde{p}}
\newcommand{\ant}[1]{{\color{blue}{#1}}}
\newcommand{\fbz}{1$^{st}$ BZ}

\title{Flocking from a quantum analogy: Spin-orbit coupling in an active fluid}
\author{Benjamin Loewe}
\affiliation{School of Physics,
Georgia Institute of Technology, Atlanta, GA, 30332, USA }
\author{Anton Souslov}
\affiliation{School of Physics,
Georgia Institute of Technology, Atlanta, GA, 30332, USA }
\affiliation{Leiden Institute of Physics,
Leiden University, Leiden, Netherlands }
\author{Paul M.~Goldbart}
\affiliation{School of Physics,
Georgia Institute of Technology, Atlanta, GA, 30332, USA }

\date{\today}

\begin{abstract}
Systems composed of strongly interacting self-propelled particles can form a spontaneously flowing polar active fluid.
The study of the connection between the microscopic dynamics of a single such particle and the macroscopic dynamics of the fluid can yield
insights into experimentally realizable active flows, but this connection is well understood in only a few select cases. We introduce a model of self-propelled particles based on an analogy with the motion of electrons that have strong spin-orbit coupling.
We find that, within our model, self-propelled particles are subject to an analog of the Heisenberg 
uncertainty principle that relates
translational and rotational noise. 
Furthermore, by coarse-graining this microscopic model, we establish expressions for the coefficients of the  Toner-Tu equations---the hydrodynamic equations that describe an active fluid composed of these ``active spins.''
The connection between self-propelled particles and quantum spins may help realize exotic phases of matter using active fluids via analogies with systems composed of strongly correlated electrons.
\end{abstract}

\maketitle
Active liquids exhibit striking phenomena due to the unusual nature of their hydrodynamics~\cite{Marchetti2013}. 
Such phenomena have been observed in naturally occurring collections of live animals~\cite{Buhl2006, Ballerini2008, Cavagna:2014hd} 
and cells~\cite{Kemkemer2000, Dombrowski2004, Szabo2006, Rafai2010,SerraPicamal2012}, as well as synthetically prepared systems of granules~\cite{Narayan2007,Kudrolli2008}, robots~\cite{Giomi2013}, colloids~\cite{Paxton2004, Palacci2010, Bricard2013}, and molecules~\cite{Surrey2001, Bendix2008, Schaller2010, Dogic2013}. 
Coarse-grained descriptions that capture these phenomena may be either constructed based solely on symmetry and lengthscale considerations or derived from simple particle-based models~\cite{Bertin2006, Bertin2009, Baskaran2010, Bricard2013}. A crucial advantage of the latter, microscopic, approach is that it connects the hydrodynamic coefficients (such as viscosity, diffusivity, and compressibility) to the microscopic parameters of the model. In experimental realizations of active fluids, this connection between microscopics and hydrodynamics can be used to construct design principles targeting the realization of novel materials and devices. For example, recent work has focused on the robustness of active liquids against disorder~\cite{Morin2016}, the design of flow patterns in confined active 
fluids~\cite{Wioland2013, Brotto2013, Pearce2015, Bricard2015, Wioland2016}, and the use of such channel networks for the design of topological metamaterials~\cite{Souslov2016} and logic gates~\cite{Woodhouse2016}.

One specific challenge is that the coarse graining of a microscopic model of self-propelled particles is, in general, technically difficult. As a result, specific counter-intuitive phenomena associated with active-liquid hydrodynamics are difficult to describe in generic, model-independent terms. In pursuit of this goal, the introduction of additional minimal models of self-propelled particles, along with their coarse-grained hydrodynamics, can serve to strengthen the connection between small- and large-scale phenomena in active systems.

\begin{figure}[ht]
\includegraphics[angle=0]{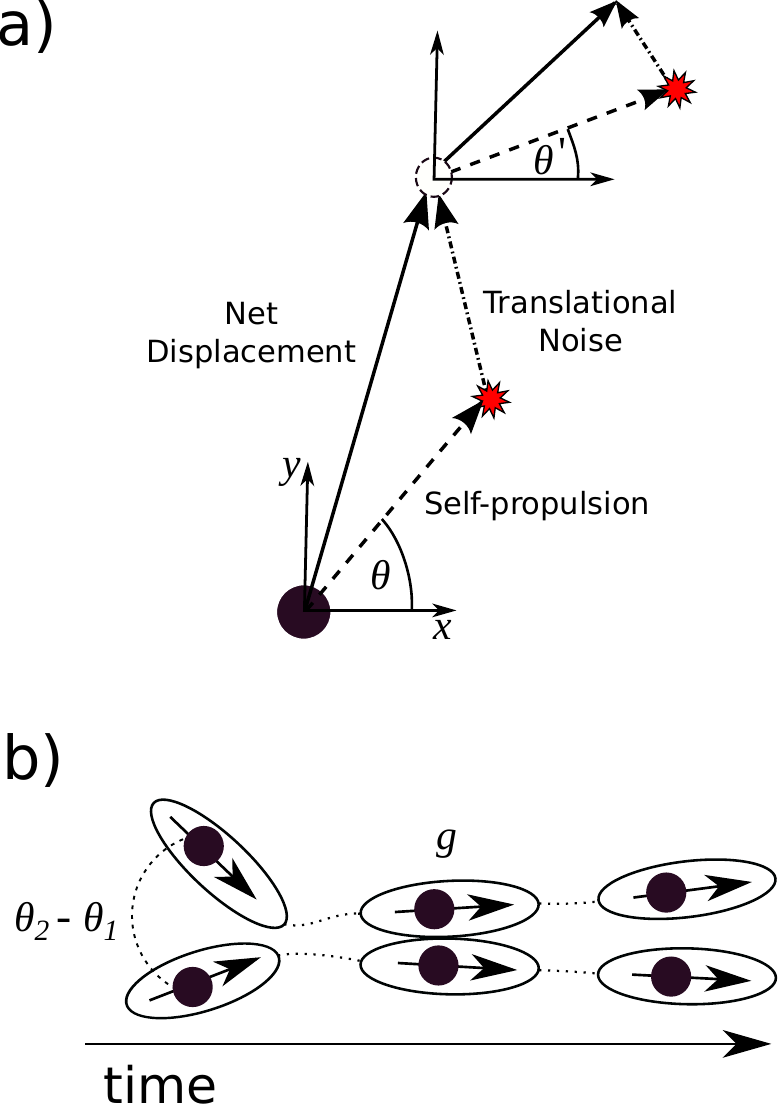}
\caption{a) Schematic illustration of single-particle dynamics in the model we consider. The particle trajectory is composed of displacements due to self-propulsion combined with translational noise. Such noise may arise, e.g., from a fluid in which the particles are suspended. In addition to the translational noise, which alters the particle displacement, 
the particle is subject to orientational noise, which alters the direction of self-propulsion. In the model we consider, the two sources of noise are intimately coupled, which leads to a relation analogous to the Heisenberg uncertainty principle.
b) For a system of many particles, we consider processes that can align the self-propulsion directions of two colliding particles. For example, elongated active particles may prefer to align due to the dynamics of their collisions.
In the coarse-grained model, we capture the strength of the alignment interaction via the parameter $g$.}
\label{dynamicaldiagram}
\end{figure}

In this work, we introduce a particularly simple, minimal model of self-propelled particles, and we explore their individual and collective statistical dynamics in order to arrive at a hydrodynamic description. The simplicity of the model stems from its connection to the Schr{\"o}dinger equation describing a quantum particle. As such, we use basic results from quantum mechanics to develop physical intuition for active-fluid phenomena. For example, we describe active-fluid analogs of such well-known quantum-mechanical concepts as spin, spin-orbit coupling, and the Heisenberg uncertainty principle.  
We discuss how the analog of a spinor can be used to introduce a propulsion direction via spin-orbit coupling. We then construct a probabilistic, Fokker-Planck interpretation for the dynamics of a single self-propelled particle in the presence of translational noise; see Fig.~\ref{dynamicaldiagram}. We show that the microscopic model we consider includes feedback between rotational and translational noise, which we interpret as an analog of the Heisenberg uncertainty relation.  Crucially, we use this single-particle model to construct a hydrodynamic description of a system of many self-propelled particles. We thus obtain simple relations between the coefficients in the Toner-Tu model~\cite{Toner1995} and the microscopic parameters of the individual particles under consideration, including their interactions. We are then able to conclude that, as for any model in the Toner-Tu universality class, the many-particle system we consider exhibits long-range orientational order in two dimensions as a consequence of activity~\cite{Toner1995}.

\section{Model of Self-Propelled Particles}
We begin with the well-known connection between non-relativistic quantum mechanics (described via the Schr{\"o}dinger equation) and classical statistical mechanics (described via the diffusion equation). 
Consider the Schr\"{o}dinger equation: 
\begin{equation}
i \hbar \partial_{t}\Psi= \hat{H} \Psi.
\end{equation}
For the free-particle Hamiltonian operator $\hat{H} = \hat{p}^2/2m = - \hbar^2 \nabla^2 /2 m $ (in the position representation), a rotation of time into the imaginary axis via $t \rightarrow - i t$ transforms this Schr{\"o}dinger equation into the diffusion equation:
\begin{equation}
\partial_t \Psi = \frac{\hbar}{2 m} \nabla^2\Psi.
\end{equation}
 In the diffusion equation, $\Psi$ can be identified with the particle density $\rho$ and $\frac{\hbar}{2 m}$ with the diffusion constant $D$. This bridge allows us to use  tools from quantum mechanics to characterize classical stochastic phenomena. However, this approach does not capture self-propulsion or spontaneous active flow, which cannot be described via the diffusion equation. 

In order to capture self-propulsion, each particle ought to carry information about its direction of motion, for which we need to introduce additional degrees of freedom. 
On the quantum side of the analogy, these degrees of freedom capture the quantum spin state. 
For a system having spin, $\Psi$ is an $n$-component spinor (we consider $n=2$),
and has additional symmetries with respect to spin rotation, which we consider in the following subsection. Significantly, we consider a two-dimensional quantum system with \emph{spin-orbit coupling}, i.e., particles whose momentum operator is coupled to their spin state. The Hamiltonian (with $\hbar = 1$ henceforth) of the system is:
\begin{equation}
\hat{{\cal H}} =  \frac{1}{2}\left[ \bm{\sigma}\cdot \nabla+m (I-\sigma_z)-\frac{1}{\kappa}\nabla^2   \right],
\label{hamiltonian}
\end{equation}
where $\bm{\sigma}\cdot \nabla \equiv\sigma_x \partial_x + \sigma_y \partial_y$ is the spin-orbit coupling term and
\begin{equation}
\sigma_x \equiv \left(\begin{array}{cc}
    0&1 \\
    1&0 \end{array}\right), \,
\sigma_y \equiv \left(\begin{array}{cc}
    0&-i \\
    i&0 \end{array}\right),
\, \sigma_z \equiv \left(\begin{array}{cc}
    1&\phantom{-}0 \\
    0&-1 \end{array}\right) \nonumber
\end{equation}
are the Pauli spin matrices. Of these, only ($\sigma_x$, $\sigma_y$) are associated with the two-dimensional coordinate frame. In the quantum system, if $\kappa \rightarrow \infty$ then $\hat{{\cal H}}$ becomes the two-dimensional Dirac Hamiltonian in the Weyl representation (up to a global energy shift). 

Dimensionality plays an important role in this description: the one-dimensional Dirac equation is given by $\hat{{\cal H}}$ with $\kappa \rightarrow \infty$ and for $\Psi$ independent of $y$. Rotated into imaginary time, the one-dimensional case yields a description of the time-dependent probability distribution of a persistent random walker. The stochastic process corresponding to such a walker is a Poisson process, and the one-dimensional Dirac equation in imaginary time can be restated as the telegrapher's equations describing this process. Such walkers move along a line at constant speed and with a turning rate (i.e., a rate for changing direction) given by $m$. 
This analogy has its origins in the path-intergral formulation of the Dirac equation~\cite{FeynmanHibbs}, also restated in imaginary time in Refs.~\cite{Schulman1,Schulman2,Schulman3}.

Although we develop some intuition by considering the quantum side of the analogy, we mostly focus on the mathematical description of self-propelled particles by performing a rotation of time into the imaginary axis: $t \rightarrow i t$. One of our main conclusions is that the (imaginary-time Schr{\"o}dinger) equation in two dimensions,
\begin{equation}
- \partial_t \Psi = \hat{{\cal H}} \Psi,
\label{eq:dynamic}
\end{equation}
with $\hat{{\cal H}}$ given by Eq.~(\ref{hamiltonian}), describes the time-evolution of the probability distribution $\Psi(x,t)$ of a self-propelled particle subject to two sources of noise: translational noise (controlled by the strength of the diffusion constant $1/\kappa$),
and rotational noise in the orientation angle [controlled by the parameter $m$ in the term $m (I-\sigma_z)t$]. This latter term describes the ability of active particles to change their direction of motion, as it does for particles in the one-dimensional Dirac equation~\cite{Schulman1,Schulman2,Schulman3}. We also show that, unlike in the one-dimensional case, the two-dimensional Dirac Hamiltonian does not consistently describe the probability density of a single, self-propelled particle:
to ensure a physical description, $\kappa$ must be constrained to be less than $8m$. We now proceed to derive these results.

\subsection{Spin, rotation, and velocity}
Let us now discuss the motivation for using spin as the carrier of information about the direction of propulsion. We use the spinorial representation of the rotation group, in which a rotation by angle $\theta$ around the axis $\hat{\textbf{n}}$ is generated by the unitary operator $\hat{U}_{\hat{\bm{n}},\theta}\equiv\exp(-i \theta \hat{\textbf{n}}\cdot\bm{\sigma}_3/2)$, and where we have used the spin vector $\bm{\sigma}_3$ defined by $\bm{\sigma}_3 \equiv (\sigma_x,\sigma_y,\sigma_z)$.
We are considering particles constrained to be in the $xy$-plane and, therefore, all rotations are around the $z$-axis: $\sigma_z$ generates this abelian rotation group. These rotation operators are given by
\begin{equation}
\hat{U}_{\theta}=\exp(-i \theta \sigma_z/2)=e^{-i \theta/2}\left(\begin{array}{cc}
    1&0 \\
    0&e^{i \theta} \end{array}\right).
\end{equation}
As the global phase $e^{-i \theta/2}$ does not change the physical quantum state, we redefine the operator as
\begin{equation}
\hat{U}^\prime_{\theta}=\left(\begin{array}{cc}
    1&0 \\
    0&e^{i \theta} \end{array}\right).
\end{equation}

The action of the rotation $\hat{U}^\prime_{\theta}$ on the spinor $(a,b)$ transforms it into the spinor $(a,b e^{i \theta})$.  Note that the second component is rotated in the complex plane by the angle $\theta$. Thus, the phase of this spinor component can be interpreted as the orientation of a polar particle, i.e., a particle that carries information about its orientation. Without loss of generality, we choose a global phase such that the first component of the spinor is real. Then, the spinor describing a particle oriented along $\hat{\bm{n}}=(\cos\theta,\sin\theta)$ is given by
\begin{equation}
\xi=\left(\begin{array}{cc}
    s_1 \\
    s_2 = |s_2|e^{i \theta} \end{array}\right),
\end{equation}
with $s_1$ real. The particle orientation is then given in terms of the real ($\Re$) and imaginary ($\Im$) parts of $s_2$  by $\hat{\bm{n}}\equiv(\Re s_2 ,\Im s_2)/|s_2|$.

We use spinors to describe the \emph{stochastic} nature of motion and orientation for a self-propelled particle. 
We aim to capture quantities characterizing this particle such as the (scalar) probability $\rho(\bm{r})$ of finding the particle at position $\bm{r}$ (regardless of its orientation) and corresponding probability current $\bm{j}(\bm{r})$. We show that the spinor encodes this information via $\rho = \Re(s_1)$ and $\bm{j}= (\Re s_2 ,\Im s_2)$. We can then construct the probability density $P(\bm{r},\theta)$ for finding the particle at position $\bm{r}$ and oriented along angle $\theta$ via
\begin{equation}\label{eq:p}
P(\bm{r},\theta)=s_1+\bm{v}(\theta)\cdot \bm{s}_2,
\end{equation}
where $\bm{v}(\theta) \equiv (\cos \theta,\sin\theta)$ and $\bm{s}_2=(\Re s_2 ,\Im s_2)$.

For a self-propelled particle, motion and orientation are coupled. Using the probability density in Eq.~(\ref{eq:p}), we note that the average of the particle's orientation at position $\bm{r}$ is proportional to $\bm{j}$. We can conclude that the eigenvectors of the $x$- and $y$-components of the operator $\bm{\sigma} = (\sigma_x,\sigma_y)$ correspond to particles oriented along the $x$- and $y$-directions, respectively. We then note that $\bm{\sigma}\cdot \nabla$ is reminiscent of the convective derivative $\bm{v}\cdot\nabla$: $\bm{\sigma}\cdot \nabla$ convects the probability density in the direction along which the spinor points.
Alternatively, this relation between $\bm{\sigma}$ and the velocity operator can be gathered directly from the Heisenberg equation of motion. If we consider the simplified Hamiltonian $\hat{H} = \bm{\sigma}\cdot \nabla$, 
we can express the time-derivative of the position operator $\bm{r}$ (i.e., the velocity operator) as
\begin{equation}
\frac{d\bm{r}}{dt}=[\hat{H},\bm{r}]=\bm{\sigma}\cdot[\nabla,\bm{r}]=\bm{\sigma}.
\end{equation}
As a result of the additional terms in the Hamiltonian~(\ref{hamiltonian}) of the model we consider, the translational noise contributes along with  $\bm{\sigma}$ to the velocity operator.

\subsection{Probabilistic interpretation}
We now proceed to demonstrate the link between (imaginary-time Schr{\"o}dinger) Eq.~(\ref{eq:dynamic}) and the probability densities and currents of self-propelled particles. To do so, we decompose the spinor $\Psi$ into real and imaginary parts:
\begin{equation}
\Psi=\left(\begin{array}{cc}
    \rho + i \chi \\
    j_x + i j_y \end{array}\right),
\end{equation}
where $\rho$, $\chi$, $j_x$, and $j_y$ are real-valued functions of the position $\bm{r}$ and time $t$ (and are independent of $\theta$). With this parametrization, Eq.~(\ref{eq:dynamic}) becomes
\begin{align}
    -2\partial_t\rho &= \nabla \cdot \bm{j} -\frac{1}{\kappa}\nabla^2 \rho \nonumber, \\
    -2\partial_t\chi&=-\nabla_{\perp}\cdot \bm{j}-\frac{1}{\kappa}\nabla^2 \chi \nonumber, \\
		-2\partial_t \bm{j}&=\nabla \rho - \nabla_{\perp}\chi+2 m \bm{j} -\frac{1}{\kappa}\nabla^2 \bm{j},
		\label{system}
\end{align}
where we have introduced $\bm{j} \equiv (j_x, j_y)$ and, for any vector $\bm{a}$, $\bm{a}_{\perp}\equiv(a_y,-a_x)$. 
 Note that the  first of these equations can be interpreted as a continuity equation, with $\rho$ taking the role of a density and with both $\bm{j}$ and $\nabla \rho$ contributing to the current. Furthermore, we can interpret $\chi$ as a gauge degree of freedom for the orientation of the local coordinate frame. We make the simplest choice of gauge: $\chi=0$. Substituting this condition into Eqs.~(\ref{system}), we find
\begin{align}
-2\partial_t\rho &= \nabla \cdot \bm{j} -\frac{1}{\kappa}\nabla^2 \rho, \label{consistent1} \\
\nabla_{\perp}\cdot \bm{j} &=0, \label{consistent2}\\
-2\partial_t \bm{j}& =\nabla \rho +2 m \bm{j} -\frac{1}{\kappa}\nabla^2 \bm{j}. \label{consistent3}
\end{align}
We check the consistency of our gauge choice by noting that if the initial conditions satisfy Eq.~(\ref{consistent2}), the evolution given by Eqs.~(\ref{consistent1}, \ref{consistent3}) remains consistent with Eq.~(\ref{consistent2}). Indeed, we find this consistency condition to hold by applying $\nabla_{\perp}$ to Eq.~(\ref{consistent3}):
\begin{equation}
-2\partial_t (\nabla_{\perp}\cdot \bm{j})= 2 m (\nabla_{\perp}\cdot\bm{j})-\frac{1}{\kappa}\nabla^2 (\nabla_{\perp}\cdot\bm{j}).
\label{consistent4}
\end{equation} 

In what follows, we parametrize the velocity of self-propulsion via $\bm{v}(\theta)=(\cos\theta,\sin\theta)$, and use Eq.~(\ref{consistent1}) to show that the system of equations~(\ref{consistent1}-\ref{consistent3}) is equivalent to a Fokker-Planck equation that describes the dynamics of the probability density $P(\bm{r}, \theta) \equiv \rho + \bm{v}(\theta)\cdot \bm{j}$. 
Physically, $P$ describes the probability of having a particle near $\bm{r}$ and oriented at an angle near $\theta$. 
We first decompose the current into components parallel to and perpendicular to the velocity $\bm{v}$ via ${\bm{j}=(\bm{v}\cdot\bm{j})\bm{v}+(\bm{v}_{\perp}\cdot\bm{j})\bm{v}_{\perp}}$
and substitute this identity into the continuity equation~(\ref{consistent1}).
To get the dynamics of the distribution of the orientation angle $\theta$, we multiply Eq.~(\ref{consistent3}) by $\bm{v}$. We add this equation for the time evolution of the current to the continuity equation to find
an equation for the probability density $P$:
\begin{equation}
-2\partial_t P= (\bm{v}\cdot\nabla) P+(\bm{v}_{\perp}\cdot\nabla)(\bm{v}_{\perp}\cdot\bm{j})+2 m(\bm{v}\cdot\bm{j})-\frac{1}{\kappa} \nabla^2 P.
\label{midstep}
\end{equation}
The two terms that are not yet expressed in terms of $P$ can be addressed in the following way. 
First, note that $\bm{v}\cdot\bm{j}$ encodes orientational diffusion: $\partial^2_{\theta} \bm{v}=-\bm{v}$, and thus
\begin{equation}
\bm{v}\cdot\bm{j}=-\partial^2_{\theta}(\bm{v}\cdot\bm{j})=-\partial^2_{\theta}P.
\label{midstep2}
\end{equation}
We also find an extra component to diffusion that couples translational noise, rotational noise, and convection. 
This can be obtained using the identity $\bm{v}_{\perp}\cdot\bm{j}=-\partial_{\theta}P$:
\begin{equation}
(\bm{v}_{\perp}\cdot\nabla)(\bm{v}_{\perp}\cdot\bm{j}) = (\bm{v}\cdot\nabla) P-\partial_{\theta} \partial_x(P \sin\theta)+\partial_{\theta} \partial_y(P \cos\theta).
\label{midstep3}
\end{equation}
By substituting all of the diffusive and convective terms in Eqs.~(\ref{midstep2}-\ref{midstep3}) into Eq.~(\ref{midstep}), we arrive at the Fokker-Planck equation for $P$:
\begin{widetext}
\begin{equation}
\partial_t P= -(\bm{v}\cdot\nabla) P +\frac{1}{2}\left[\partial_{\theta}\,\partial_x(P \sin\theta)+\partial_{\theta}\,\partial_y(-P \cos\theta)+2 m\,\partial^2_{\theta}P+\frac{1}{\kappa}\nabla^2 P \right].
\label{Fokker}
\end{equation}
\end{widetext}
From this formulation, we may note that $m$ plays the role of a diffusion constant for the orientation $\theta$, and $\kappa^{-1}$ plays that role for the position $\bm{r}$.
The derivation of Eq.~(\ref{Fokker}) is one of our main results: we showed that the Fokker-Planck equation~(\ref{Fokker}) is equivalent to the imaginary-time Schr\"{o}dinger equation~(\ref{eq:dynamic}) with the Hamiltonian~(\ref{hamiltonian}), which includes a spin-orbit coupling term.

\subsection{Microscopic Langevin equation}

We now derive the precise microscopic model that corresponds to the Fokker-Planck equation~(\ref{Fokker}).
Before doing so, first note that for a Fokker-Planck equation to represent a stochastic microscopic model, the associated diffusion matrix must be positive-definite. In this subsection, we derive and analyze the drift vector and diffusion matrix for the model defined by Eqs. (\ref{hamiltonian}-\ref{eq:dynamic}), and derive the conditions under which an underlying microscopic model exists.
To begin this analysis, we first define $z \equiv a \theta$, where $a$ is a particle lengthscale, and we re-express $\theta$ in terms of $z$ to ensure that the different entries in the diffusion matrix have the same dimensionality.
We compare Eq.~(\ref{Fokker}) with the usual Fokker-Planck equation, viz.,
\begin{equation}
\partial_t P= -\sum_{i=1}^3\partial_i(\mu_i P)+\frac{1}{2}\sum_{i,j=1}^3\partial_{i} \partial_{j}(D_{ij}P),
\end{equation}
in which $\bm{\mu}$ is the drift vector and $D$ is the diffusion matrix. We then read off as follows:
\begin{equation}
\bm{\mu}=(\bm{v},0)=(\cos\theta,\sin\theta,0),
\end{equation}
using a vector notation in which the third component corresponds to $\theta$ and
\begin{equation}
D=\left(\begin{array}{c c c}
   1/\kappa & 0 & (a/2)\sin\theta\\
		0 & 1/\kappa & -(a/2)\cos\theta\\
    (a/2)\sin\theta & -(a/2)\cos\theta & 2m a^2 \\
	\end{array}\right).
\label{diff_tensor}
\end{equation} 
As mentioned above, in order to construct a particle-based model for Eq. (\ref{Fokker}), it is necessary that $D$ be positive-definite. To check this, we examine its eigenvalues $(\lambda_1,\,\lambda_{+},\,\lambda_{-})$: 
\begin{align}
\lambda_{1}&=1/\kappa,\\
\lambda_{\pm}&=\frac{1}{2}\left(2 m a^2+\frac{1}{\kappa}\right) \pm \left[(2m a^2-1/\kappa)^2+a^2\right]^{-1/2}.
\end{align}
$\kappa$, $a$, and $m$ real and positive guarantees that $\lambda_{1}$ and $\lambda_+$ are positive.
From the expression for $\lambda_{-}$ we conclude that $D$ is positive-definite if and only if
\begin{equation}
8 m>\kappa.\label{eq:cond}
\end{equation}
Thus, in order for Eq.~(\ref{Fokker}) with $m > 0$ to correspond to a microscopic model, translational noise must be present, because without translational noise (i.e., in the limit $\kappa \rightarrow \infty$) there isn't an orientational noise parameter $m$ that satisfies Eq.~(\ref{eq:cond}). This implies that, in two dimensions, the Dirac equation alone cannot describe a self-propelled particle, in contrast to the one-dimensional case.	
As mentioned in the introduction to this section, the one-dimensional (imaginary-time) Dirac equation is equivalent to the telegrapher's equations, which do describe a stochastic process that can be interpreted as self-propulsion in one dimension.

From Eq.~(\ref{Fokker}), further insight into the relationship between $m$ and $\kappa$ and their physical interpretations can be gained by deriving the connection between this Fokker-Planck equation and the underlying microscopic process, i.e., the stochastic Langevin equation
\begin{equation}
d\bm{R}_t=\bm{\mu}(\bm{R}_t,t)dt+\Sigma(\bm{R}_t,t)d\bm{W}_t.
\label{General_Langevin}
\end{equation}
In Eq.~(\ref{General_Langevin}), $\bm{R}_t$ has $N$ components (corresponding to the random variables), $\bm{\mu}$ is an $N$-component associated drift vector, $\Sigma(\bm{R}_t,t)$ is an $N\times M$ matrix, and $\bm{W}_t$ is an $M$-dimensional Wiener process interpreted in either the It\^{o} or Stratonovich sense~\cite{Gardiner, Kampen}. For example, in the present case of $N=3$, $\bm{R}_t$ consists of the position vector $\bm{R}=(x,y)$ and orientation angle $\theta$.

Formally, this interpretation can be established for an It\^{o} process by considering a diffusion matrix $D$ of the form $\Sigma \Sigma^T$. Notice that if one such $\Sigma_0$ furnishes this decomposition then, for any orthogonal matrix $R$, $\Sigma^\prime = \Sigma_0\,R$ also satisfies it, so there are many different Langevin equations that yield the same Fokker-Planck equation.  By using a Cholesky decomposition \cite{Golub2013}, we find a particular solution for the case $M=3$:
\begin{equation}
\Sigma= \left(\begin{array}{c c c}
 \kappa^{-1/2} & 0 & 0 \\
 0 & \kappa^{-1/2} & 0 \\
 \frac{1}{2}\,a\kappa^{1/2} \sin\theta & -\frac{1}{2}\,a \kappa^{1/2} \cos\theta & \sqrt{2}\,a\left[  m  - \frac{\kappa}{8}\right]^{1/2} 
\end{array}\right),\nonumber
\end{equation}
which yields the following Langevin equation:
\begin{align}\label{langevin1}
d\bm{R} & =\bm{v}(\theta)\,dt+\kappa^{-1/2}\bm{\xi}\,dt,\\
d\theta & = \frac{1}{2} \kappa^{1/2}(\bm{v}_{\perp}(\theta)\cdot\bm{\xi})\,dt+\sqrt{2}\left[m-\frac{1}{8} \kappa\right]^{1/2} \xi_3\,dt.
\label{langevin2}
\end{align}
Here, $\bm{\xi}=(\xi_1,\xi_2)$ is the two-dimensional translational noise that acts on the position of the particle, whereas $\xi_3$ is a rotational noise influencing the polarization angle $\theta$. Note that to interpret this microscopic model we have assumed that Eqs.~(\ref{langevin1}, \ref{langevin2}) are It\^{o} stochastic differential equations. Generally, this differs from a Stratonovich process by an extra, noise-induced, drift vector having components
\begin{equation}
\mu_i= \frac{1}{2}\sum_{k,j=1}^3(\partial_j \Sigma_{ik})\Sigma_{jk}.
\end{equation}
However, in the case we are considering, the corresponding term is identically zero, and thus Eqs.~(\ref{langevin1}-\ref{langevin2}) can also be seen as a Stratonovich stochastic differential equation.

\subsection{Noise and the uncertainty principle}
Let us now discuss the physical picture of the single-particles dynamics described by Eqs.~(\ref{langevin1}-\ref{langevin2}).
This microscopic model has similarities to the models of active particles used, e.g., in Refs.~\cite{Vicsek, Marchetti2013}. 
At each instant in time, a particle is oriented at an angle $\theta$ and attempts to propagate in this direction at a constant speed. 
However, translational noise can change the direction of propagation away from the particle polarization.
As a unique feature, the model we consider has feedback between translational and rotational noise:
the larger the translational noise, the weaker the rotational noise.
Quantitatively, if we define $\alpha_{\xi}$ to be the angle that the force from the translational noise, $\bm{\xi}=(\xi_1,\xi_2)$, makes with respect to the $x$-axis, we can rewrite the rotational noise term $\kappa^{1/2}(\bm{v}_{\perp}(\theta)\cdot\bm{\xi})/2$ in Eq.~(\ref{langevin2}) as
\begin{equation}
\frac{1}{2}\kappa^{1/2}|\bm{\xi}|\sin(\theta-\alpha_{\xi}).
\label{eq:noise}
\end{equation}
From Eq.~(\ref{eq:noise}), we observe that 
particles in effect try to oppose the translational noise, and prefer to align opposite to the direction of each kick. 
This coupling acts as a guidance system: in the absence of translational noise, the particle does not know which way to point. This is a consequence of how $\kappa$ enters Eq.~(\ref{langevin2}): 
the translational noise strength is \emph{inversely} proportional $\kappa$, whereas the rotational noise strength is proportional to $\kappa$. Thus, the particle depends on feedback from translational noise to decide where to go. 
(A curious analogy emerges from the physics of hair cells in the inner ear, which depend on the presence of external noise to complete their function~\cite{haircells}.)

To further examine this feedback feature, consider the extreme case in which $m$ is only slightly bigger than the lower-bound of $\kappa/8$, i.e., $m = \kappa/8+\delta$, with $\delta/\kappa \ll 1$.  If $\delta$ is sufficiently small then rotational noise becomes irrelevant, compared to the large translational noise, and Eq.~(\ref{langevin2}) becomes
\begin{equation}
d\theta = \frac{1}{2}\kappa^{1/2}|\bm{\xi}|\,\sin(\theta-\alpha_{\xi})\, dt.
\end{equation}
In this regime, the noise dominates over the self-propelled aspect of particle motion. The strength of this noise, quantified by $\kappa^{-1}$, is not subject to any restrictions, and both large- and small-noise regimes are physically accessible.
 
Curiously, the interplay between the strength of the translational noise and its effect on the polarization is an expression of an uncertainty principle in this model. To see this, disregard the drift, and consider the feedback on the angle as simple additive noise. One then finds $\langle[\bm{R}(t)-\bm{R}(0)]^2\rangle\sim 4t/\kappa$ and $\langle[\theta(t)-\theta(0)]^2\rangle \sim t \kappa/2$, which suggest the relation:
\begin{equation}
\frac{1}{t^2}\langle[\bm{R}(t)-\bm{R}(0)]^2\rangle\left\langle[\theta(t)-\theta(0)]^2\right\rangle \sim 2.\label{eq:hup}
\end{equation}
This is a direct analog of the Heisenberg uncertainty principle, which relates the uncertainty of the position and velocity 
(here captured by orientation $\theta$) of a quantum particle.

We now compare this microscopic model with others discussed in the literature. One related example involves a system composed of self-propelled hard rods that also experience translational noise, as examined in Refs.~\cite{HardRods1,HardRods2,HardRods3}. 
In these models, the translational and orientational noises are assumed to be uncorrelated. A situation closer to ours is explored in Ref.~\cite{Romanczuk2012}, in which the translational noise affects both orientational and spatial diffusion. 
In that case, the effects of these correlations have been examined in the inertial regime, in which Fokker-Planck dynamics are not equivalent to the imaginary-time
Schr\"{o}dinger equation that we examine here.

To conclude this section, let us generalize this interplay between translational and rotational noise and give it an arbitrary strength. In this case, Eq.~(\ref{langevin2}) acquires an additional arbitrary (real) parameter $\lambda$ via:
\begin{equation}
 d\theta =\frac{1}{2}\kappa^{1/2}\lambda|\bm{\xi}|\sin(\theta-\alpha_{\xi})\,dt+\sqrt{2}\left[m-\frac{\kappa\lambda^2}{8}\right]^{1/2} \xi_3\,dt \nonumber
\end{equation}
(with $m>\kappa\lambda^2 /8$). This parameter $\lambda$ controls the response of a self-propelled particle to translational noise. The sign of $\lambda$ determines the type of response: for $\lambda < 0$, the particle turns in the direction of any translational kick, whereas for $\lambda > 0$, as in the case above,  the particle reacts in opposition to the kick. The Fokker-Planck equation associated with the Langevin dynamics of Eq.~(\ref{langevin1}) is given by
\begin{align}
\partial_t P &= -(\bm{v}\cdot\nabla) P +\frac{1}{2}\lambda\left[\partial_{\theta} \partial_x(P \sin\theta)-\partial_{\theta} \partial_y (P \cos\theta)
\right] \nonumber\\
&+ m \partial^2_{\theta}P+\frac{1}{2\kappa}\nabla^2 P .
\label{lambda_fokker}
\end{align}
Although the additional parameter $\lambda$ introduces more flexibility into the model, it destroys the uncertainty principle~(\ref{eq:hup}) and the bridge with the Schr\"{o}dinger equation describing a two-component spinor.

\section{Hydrodynamics of Active Spins}
In the previous section, we concluded that the Schr\"{o}dinger equation for a model with a spin-orbit coupling term can be interpreted as an equation for the probability density of a self-propelled particle. In this section, we start with the \textit{many-body} version of such a model, and go on to derive the coarse-grained hydrodynamic description of this polar active fluid.

In the noninteracting limit, the $N$-body Schr\"{o}dinger equation can be written in terms of the elementary generalization $\hat{{\cal H}}_N$ of the one-body Hamiltonian~(\ref{hamiltonian}), given by
\begin{equation}
\hat{{\cal H}}_N = \frac{1}{2}\sum_{i=1}^N\left[\sigma_i\cdot \nabla_i+m (I_i-\sigma_{z,i})-\frac{1}{\kappa}\nabla_i^2   \right],
\label{NHamiltonian}
\end{equation}
where the summation $i=1,\ldots,N$ is performed over the $N$ particles.

For noninteracting particles, we extract the probabilistic interpretation of the many-body wavefunction by noting that the probability density for many independent processes must obey $P_N(\bm{r}_1,\theta_1;\ldots;\bm{r}_N,\theta_N)=\prod_{i=1}^N P_i(\bm{r_i},\theta_i)$.
Therefore, the one-particle quantities $\rho$ and $\bm{j}$ can be written as
\begin{align}
\rho_i(\bm{r}_i) &=\frac{1}{2 \pi}\int_0^{2\pi} d\theta_i \, P_i(\bm{r}_i,\theta_i),\\
\bm{j}_i(\bm{r}_i) &= \frac{1}{\pi}\int_0^{2\pi}d\theta_i \, \bm{v}(\theta_i) \, P_i(\bm{r}_i,\theta_i).
\end{align}
On the other hand, the many-body spinor associated to $P_N$ has $2^N$ components and the following structure: $\Psi^N_{\sigma_1,\ldots,\sigma_N}(\bm{r}_1,\ldots,\bm{r}_N)=\prod_{i=1}^N \Psi_{\sigma_i}(\bm{r}_i)$, as in the case of many non-interacting and uncorrelated quantum particles. Notice that the probability density $P$ can capture more details regarding the distribution of the angular variable $\theta$ than the components of the spinor $\Psi$ can encode. Indeed, the spinorial description assumes that only the first two Fourier modes in the angle $\theta$ are relevant, and disregards all higher Fourier components. Thus, in order to reduce a description in terms of $P$ to one in terms of $\Psi$, the quantities $\int_0^{2\pi}\! e^{i n \theta} P(\theta) d\theta $ must be negligible for all $|n|\geq 2$.

Before we go on to include many-particle interactions, we generalize the route that took us from the Schr\"{o}dinger equation~(\ref{eq:dynamic}) to the Fokker-Planck equation~(\ref{Fokker}) to include multiple particles. We then arrive at the following many-particle Fokker-Planck equation:
\begin{align}
\partial_t P_N &= -\sum_{i=1}^N(\bm{v}_i\cdot\nabla_i) P_N \nonumber\\
&+\frac{1}{2}\sum_{i=1}^N \big[\partial_{\theta_i} \partial_{x_i} (P_N \sin\theta_i)-\partial_{\theta_i} \partial_{y_i} (P_N \cos\theta_i)
 \nonumber\\
&+2 m \partial^2_{\theta_i}P_N+\frac{1}{\kappa}\nabla_i^2 P_N \big].
\label{freeFokk}
\end{align}
In order to uncover collective phenomena, we need to consider inter-particle interactions. For example, let us consider the alignment interaction typical of, e.g., the XY model, which can be included via a potential in the many-particle Langevin equation~(\ref{langevin2}):
\begin{align}
 d\theta_i = & V_i(\{\bm{r},\theta\})\,dt+ \frac{1}{2}\kappa^{1/2}(\bm{v}_{\perp}(\theta)\cdot\bm{\xi})\,dt \nonumber \\
 & +\sqrt{2}\left[m-\frac{\kappa}{8}\right]^{1/2}\xi_3 \, dt,\label{eq:dt}
\end{align}
wherein $(\{\bm{r},\theta\})$ is a shorthand notation for $(\bm{r}_1,\theta_1;\ldots;\bm{r}_N,\theta_N)$.
In Eq.~(\ref{eq:dt}), the inter-particle interactions are encoded in the potential $V_i$, which is defined via
\begin{equation}
V_i(\{\bm{r},\theta\}) \equiv g \sum_{j( \neq i)} R(\bm{r}_i-\bm{r}_j) \sin(\theta_j-\theta_i)
\end{equation}
and acts with interaction strength $g$. The inter-particle separation enters the interaction potential via the function $R(\bm{r}_i-\bm{r}_j)$, which includes the characteristic range of the interactions. We now add this two-particle interaction to the many-particle Fokker-Planck equation (\ref{freeFokk})
\begin{align}
\partial_t P_N &= -\sum_{i=1}^N\left[(\bm{v}_i\cdot\nabla_i)P_N+\partial_{\theta_i}\left(V_i(\{\bm{r},\theta\}) P_N\right)\right] \nonumber \\
& +\frac{1}{2}\sum_{i=1}^N\big[\partial_{\theta_i} \partial_{x_i} (P_N \sin\theta_i)-\partial_{\theta_i} \partial_{y_i} (P_N \cos\theta_i) \nonumber\\
&+2 m \partial^2_{\theta_i}P_N+\frac{1}{\kappa}\nabla_i^2 P_N \big].
\label{fullFokk}
\end{align}
Equation~(\ref{fullFokk}) describes the time evolution for the probability distribution of many interacting self-propelled particles.

\subsection{Self-consistent approximation}
Instead of trying to exactly solve Eq.~(\ref{fullFokk}), in the present section we introduce a self-consistent approximation. To do this, we rewrite the potential as
\begin{equation}
V_i(\{\bm{r},\theta\}) = g\,\Im\left(h(\bm{r}_i)\,e^{i\alpha(\bm{r}_i)}\,e^{-i \theta_i}\right),
\end{equation}
where
\begin{equation}
h(\bm{r}_i)\,e^{i\alpha(\bm{r}_i)} \equiv \sum_{j(\neq i)} R(\bm{r}_i-\bm{r}_j)\,e^{i\theta_j}.
\end{equation}
The defining assumption of the self-consistent approximation is that
\begin{equation}
h(\bm{r}_i)\,e^{i\alpha(\bm{r}_i)} \approx \left\langle h(\bm{r}_i)\,e^{i\alpha(\bm{r}_i)} \right\rangle_C,
\label{eq:cav}
\end{equation}
where the average $\langle \cdots \rangle_C$ is taken with respect to the conditional probability that one of the particles is at position $\bm{r}_i$ and oriented along $\theta_i$:
\begin{equation}
P(\{\bm{r}_j,\theta_j\}_{j(\neq i)} | \bm{r}_i,\theta_i)=\prod_{\ell(\neq i)}P_{\ell}(\bm{r}_{\ell},\theta_{\ell}).
\end{equation}
This approximation treats the inter-particle interaction as an external potential due to the average effect of all the other particles. An explicit computation of the conditional average in Eq.~(\ref{eq:cav}) leads to
\begin{equation}
\left\langle h(\bm{r}_i)\,e^{i\alpha(\bm{r}_i)} \right\rangle_C = \sum_{j(\neq i)}\int_0^{2\pi}d\theta_j\,e^{i\theta_j}\left\langle R(\bm{r}_i-\bm{r}_j)\right\rangle_{\theta_j},
\end{equation}
where 
\begin{equation}
\left\langle R(\bm{r}_i-\bm{r}_j)\right\rangle_{\theta_j} \equiv \int_A d^2r_j\,R(\bm{r}_i-\bm{r}_j)\,P_j(\bm{r}_j,\theta_j),
\end{equation}
and the integral is taken over the two-dimensional area $A$.
For simplicity, we now consider a purely local interaction [that is to say, taking $R(\bm{r}_i-\bm{r}_j) \rightarrow \delta(\bm{r}_i-\bm{r}_j)$], in which case
the self-consistency condition reduces to the simple form:
\begin{equation}
\left\langle h(\bm{r}_i)\,e^{i\alpha(\bm{r}_i)} \right\rangle_C = \sum_{j(\neq i)}\int_0^{2\pi} \! d\theta_j\,e^{i\theta_j} P_j(\bm{r}_i,\theta_j).
\label{tempmeanfield}
\end{equation}
Within the self-consistent approximation, all particles are identical and experience the same forcing. This forcing is, in turn, determined by considering the effect of a particle on its neighbors.
The assumption of identical particles leads to all particles having the same probability distributions for all observables. In terms of probabilities, we thus have $P_j(\bm{r},\theta)=P(\bm{r},\theta)$ for all $j$. 
By using Eq.~(\ref{tempmeanfield}) and taking the $N\gg 1$ limit, we obtain
\begin{equation}
\left\langle h(\bm{r}_i)\,e^{i\alpha(\bm{r}_i)} \right\rangle_C = N\int_0^{2\pi} d\theta\,e^{i\theta} P_j(\bm{r}_i,\theta).
\label{eq:h}
\end{equation}
Substituting Eq.~(\ref{eq:h}) into the expression for the potential, we find that the self-consistent potential has the form:
\begin{equation}
V_{SC}(\bm{r}_i,\theta_i)=g\,h(\bm{r}_i)\,\sin[\alpha(\bm{r}_i)-\theta_i].
\end{equation}
For convenience, we rewrite this expression using an external alignment field $\bm{h}(\bm{r})$, defined via
\begin{equation}
\bm{h}(\bm{r}) \equiv h(\bm{r}) \bm{v}(\alpha).
\end{equation}
In terms of $\bm{h}$, the potential has the form
\begin{equation}
V_{SC}(\bm{r}_i,\theta_i)=g\,\bm{h}_{\perp}(\bm{r}_i)\cdot \bm{v}(\theta).
\end{equation}

The self-consistent alignment field satisfies $|\bm{h}(\bm{r})|=h(\bm{r})$ and $\bm{h}(\bm{r})=\pi N \bm{j}$, i.e., it is a measure of the spontaneous alignment between the particle velocities. The advantage of using this self-consistent approximation is that it reduces the many-body Fokker-Planck equation to the one-particle non-linear equation, i.e., 
\begin{align}
 \partial_t P= & -(\bm{v}\cdot\nabla)P-g\partial_{\theta}[(\bm{h}_{\perp}\cdot \bm{v}) P] \nonumber \\
& +\frac{1}{2}\big[\partial_{\theta} \partial_{x} (P \sin\theta)-\partial_{\theta} \partial_{y} (P \cos\theta) \nonumber \\
&+2 m \partial^2_{\theta}P+\frac{1}{\kappa}\nabla^2 P \big].
\label{MFFokk}
\end{align}

In the present subsection we have restricted ourselves to considering a description of interacting, self-propelled particles in terms of the probability
density $P_N$ rather than in terms of the Schr\"{o}dinger equation. 
This is done out of necessity: the external potential term ${g\,\partial_{\theta}[(\bm{h}_{\perp}\cdot \bm{v}) P]}$ in Eq.~(\ref{MFFokk}) cannot be 
captured within the Hamiltonian~(\ref{hamiltonian}). To demonstrate this impossibility within a concrete example, let us consider a term in the Hamiltonian of the form $\bm{h}_{\perp}\cdot\bm{\sigma}$ as a possible candidate. 
Such a term presents two issues that cannot be overcome within the framework we are considering: (i) Such a term generates a nonzero value of $\chi$ (in the imaginary part of the spinor). 
This issue can be overcome if one considers a more general framework in which the space of quantum states includes four-spinors with the structure $\Psi=(\phi,\bar{\phi})$ as well as by including additional terms in the Hamiltonian~(\ref{hamiltonian}). (ii) More significantly, the external potential term in Eq.~(\ref{MFFokk}) couples the lowest two Fourier modes of the orientation to higher Fourier modes.
As a result, a description based on only the first two modes does not form a closed system of equations.
We thus conclude that, in general, the Schr\"{o}dinger equation in imaginary time with a spin-orbit coupling term describes single-particle dynamics only. 

\subsection{Onset of alignment}
Although the spinorial description works for single-particle dynamics only, 
we can use the Fokker-Planck description to examine the stability of the interacting isotropic active gas. In this subsection, we explore the onset of alignment due to inter-particle interactions. We follow the standard approach based on the dynamics of Fourier modes of the distribution of orientations 
$\theta$~\cite{Bertin2009}.
First, we expand the single-particle probability density in Fourier modes:
\begin{equation}
P(\bm{r},\theta)=\rho+\bm{j}\cdot\bm{v}+\sum_{n\geq 2}\bm{j}_n\cdot\bm{v}_n,
\end{equation} 
where $\bm{v}_n\equiv(\cos[n\theta],\sin[n\theta])$ and $\bm{j}_n$ are the vectors whose components are the distinct Fourier modes of P, i.e.,
\begin{subequations}
\begin{align}
j_{n,x}(\bm{r})&=\frac{1}{\pi}\int_0^{2\pi}d\theta\,\cos(\theta)\,P(\bm{r},\theta),\label{eq:exp1a}\\
j_{n,y}(\bm{r})&=\frac{1}{\pi}\int_0^{2\pi}d\theta\,\sin(\theta)\,P(\bm{r},\theta). 
\label{eq:exp1b}
\end{align}
\end{subequations}
This expansion is similar to the one introduced in Ref.~\cite{Bertin2009}; while the one in Ref.~\cite{Bertin2009} is a decomposition using complex numbers, the one used above is purely real. Substituting Eqs.~(\ref{eq:exp1a}, \ref{eq:exp1b}) into Eq.~(\ref{MFFokk}) and using the linear independence of the Fourier components leads to the following set of coupled equations describing the time-evolution of the $3$ lowest Fourier modes:
\begin{subequations}
\begin{widetext}
\begin{align}
 \partial_t\rho &=-\frac{1}{2} \nabla\cdot\left(\bm{j}-\frac{1}{\kappa}\nabla \rho\right),\label{equationsmotionm1}\\
 \partial_t j_x &= \left(\frac{1}{2\kappa}\nabla^2-m\right) j_x-\frac{1}{2}\partial_x \rho
 -\frac{3}{4}\nabla \cdot \bm{j}_2+g\left( h_x\rho-\frac{1}{2}\bm{h}\cdot\bm{j}_2\right),\\
 \partial_t j_y &= \left(\frac{1}{2\kappa}\nabla^2-m\right) j_y-\frac{1}{2}\partial_y \rho
 -\frac{3}{4}\nabla\cdot \bm{j}_{2\perp}+g\left( h_x\rho-\frac{1}{2}\bm{h}\cdot\bm{j}_{2 \perp}\right), \\
\partial_t j_{2,x} &= \left(\frac{1}{2\kappa}\nabla^2-4m\right) j_{2,x}-\nabla \cdot \bm{j}_3+g\bm{h}\ast\bm{j}-g\bm{h}\cdot\bm{j}_3,\\
\partial_t j_{2,y} &= \left(\frac{1}{2\kappa}\nabla^2-4m\right) j_{2,y}-\nabla \cdot \bm{j}_{3\perp}+g\bm{h}\ast\bm{j_{\perp}}-g\bm{h}\cdot\bm{j}_{3\perp},
\label{equationsmotion0}
\end{align}
\end{widetext}
\end{subequations}
where for compactness we have introduced the notation for the $\ast$ product of two vectors, defined via $\bm{a}\ast\bm{b} \equiv a_x b_x-a_y b_y$.  
From Eqs.~(\ref{equationsmotionm1}-\ref{equationsmotion0}) we explicitly see that the interaction terms (which are proportional to $g$) couple the higher-order Fourier modes to the lowest  ones.
Nevertheless, notice that the dependence of $\bm{j}_2$ on $\bm{j}$ is of higher order in the nonlinearity: $\bm{h}\propto\bm{j}$, and the interaction term is quadratic in the currents. 
Thus, we may deduce whether the isotropic phase is stable by performing a linear stability analysis in which we assume that the current density $|\bm{j}|$ is small compared to the particle density $\rho$. 
Then, all higher Fourier modes, such as $\bm{j}_2$, may be neglected, and we obtain a linearized theory.
As this approach neglects all stabilizing nonlinear terms, it does not yield a description of the polar active phase---we leave that task to the following subsections.

We thus proceed with examining the stability of the isotropic phase, while neglecting all non-linear terms. The linearized equations are
\begin{subequations}
\begin{align}
\partial_t\rho &=-\frac{1}{2} \nabla\cdot\left(\bm{j}-\frac{1}{\kappa}\nabla \rho\right), \label{linearizeda} \\
\partial_t \bm{j} &= \left(\frac{1}{2\kappa}\nabla^2-m\right)\bm{j}-\frac{1}{2}\nabla \rho+\frac{g N}{2 A} \bm{j}. 
\label{linearizedb}
\end{align}
\end{subequations}
In order to study the stability of the isotropic phase within Eqs.~(\ref{linearizeda}, \ref{linearizedb}), we first look for solutions of the form $(\rho,\bm{j})=(\rho_{0},\bm{j}_0)\,e^{\lambda t}$. We take spatial Fourier transforms, which re-express the gradient terms through the wavevector $\mathbf{k} \equiv (k_x, k_y)$. These steps allow us to transform the above differential equations into an eigenvalue problem, wherein $\rho_0$ and $\bm{j}_0$ act as eigenvector components and $\lambda$ as an eigenvalue.  The stability of the solutions of this system can then be analyzed by looking at the sign of the eigenvalues for each value of $\bm{k}$.  Specifically, there are three eigenvalues associated with the right-hand side of Eqs.~(\ref{linearizeda}, \ref{linearizedb}):
\begin{subequations}
\begin{align}
\lambda_1 & =-\frac{k^2}{2\kappa}-\left(m-\frac{g N}{2 A}\right), \label{eq:lam2a}\\
\lambda_{\pm} & =-\frac{k^2}{2\kappa}-\frac{1}{2}\left(m-\frac{g N}{2 A}\right)\pm\frac{1}{2}\left[\left(m-\frac{g N}{2 A}\right)^2-k^2\right]^{1/2}
\label{eq:lam2b}
\end{align}
\end{subequations}
where $k \equiv |\mathbf{k}|$. From Eqs.~(\ref{eq:lam2a}, \ref{eq:lam2b}), we note  that for small wavenumbers (or, equivalently, long wavelengths) the eigenvalues have a real part that is negative only if $m>gN/2A$.
Defining the areal number density of the particle system via $\rho_d \equiv N/A$, we may rewrite this condition as
\begin{equation}
m > g\, \frac{\rho_d}{2}. \label{eq:cond2}
\end{equation} 
Physically, this condition corresponds to the regime in which the disordered (i.e., isotropic or non-polar) fluid phase is stable with respect to 
small fluctuations. In fact, the same condition appears in the analysis of the XY and Kuramoto models of, respectively, two-dimensional spins and synchronizing oscillators; see, e.g., Refs.~\cite{Kuramoto1, Kuramoto2}. In the special case of maximally large orientational noise (i.e.,  $m\rightarrow\kappa/8$), this condition implies that
\begin{equation}
\frac{1}{4g\rho_d}> \kappa^{-1}.
\end{equation} 
This counterintuitive result states that the translational noise must be large in order for the disordered phase to be stable. This is a special feature of the model we are considering, in which the coupling between translational and orientational noise (and the resulting ``uncertainty relation'') requires a large translational noise if the orientations of the particles are to be distributed narrowly. 

\subsection{The polar active liquid}
\begin{figure}[th]
\includegraphics[angle=0]{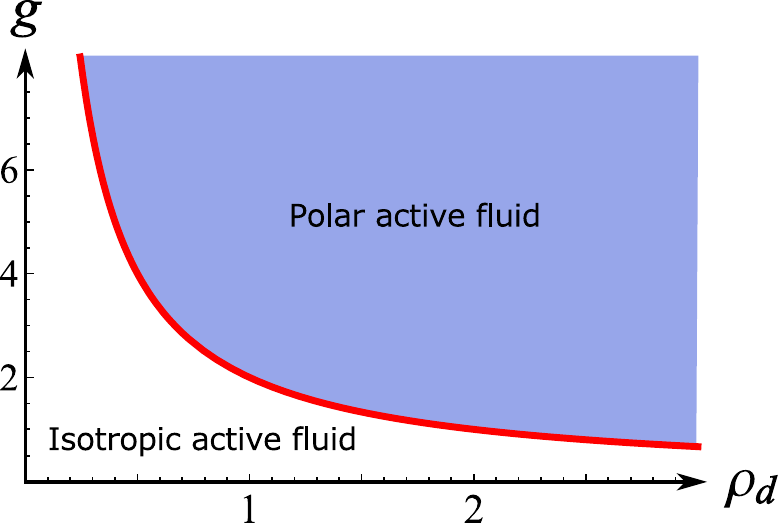}
\caption{Phase diagram of the active-spin model within the self-consistent approximation. 
Here, we take units in which $m = 1$. In this case, regions of the ordered (i.e., polar) active fluid are separated 
from regions of disordered fluid by the relation $g = 2/\rho_d$, plotted in red.
Note that the self-consistent approximation may not hold along the red transition line: strong fluctuations
may drive the transition to be discontinuous and shifted in parameter space.}
\label{fig:phasediagram}
\end{figure}

In the following subsection, we examine the behavior of long-wavelength fluctuations in the vicinity of the transition between the polar and disordered phases of the active fluid. Let us first consider the case in which the disordered phase is highly unstable, and show that the system's preferred spatially homogeneous state has rotational-symmetry-broken polar order.
As a polar fluid maintains translational symmetry, all \textit{spatial} derivatives in the Fokker-Planck equation~(\ref{MFFokk}) vanish, and the equation reduces to
\begin{equation}
\partial_t P= g\partial_{\theta}\,[(\bm{h}_{\perp}\cdot \bm{v}) P]+m\,\partial^2_{\theta}P.
\end{equation}
In this homogeneous case, pressure and translational noise are neglected, and the only obstacle to polar order is the orientational noise. 
The exact solution of this equation takes the form $P=\frac{1}{A Z} \exp\left( -\frac{g}{m} \bm{h}\cdot \bm{v} \right)$, in which $Z$ is a normalization constant. 
Although this is an exact solution, we obtain a simpler expression by only considering the case in which $|\bm{j}|$ is small.  
It is more illustrative, though, to operate in terms of a Landau theory: in this regime, we may consider only up to the first three Fourier modes, which brings us to the following systems of equations:
\begin{subequations}
\begin{align}
 \partial_t\rho &= 0 \nonumber  \\
 \partial_t j_x &= -m\,j_x+g\left( h_x\,\rho-\frac{1}{2}\bm{h}\cdot\bm{j}_2 \right),  \\
 \partial_t j_y &= -m\,j_y+g\left( h_x\,\rho-\frac{1}{2}\bm{h}\cdot\bm{j}_{2 \perp} \right),  \\
\partial_t j_{2,x} &= -4m\,j_{2,x}+g\,\bm{h}\ast\bm{j},   \\
\partial_t j_{2,y} &= -4m\,j_{2,y}+g\,\bm{h}\ast\bm{j_{\perp}}.
\end{align}
\end{subequations}
Solving these equations, and recalling that $\bm{h}=\pi N \bm{j}$, leads us to the following self-consistent equation for the alignment vector $\bm{h}$:
\begin{equation}
\bm{h}=\frac{\rho_d}{2}\left[\frac{g}{m}-\frac{1}{8}\left(\frac{g}{m}\right)^3 h^2\right] \bm{h}, \label{eq:h62}
\end{equation}
where $h=|\bm{h}|$. Solutions of Eq.~(\ref{eq:h62}) are either the trivial $\bm{h} = 0$ solution \big(which is unstable for $g\rho_d < 2 m$\big), or the solution
\begin{equation}
h = 4 \left(\frac{m^3}{\rho_d g^3}\right)^{1/2}\sqrt{\frac{\rho_d g}{2m}-1}.
\end{equation}
This solution is only physical for $g\rho_d\geq 2 m$. Thus, this inequality presents the boundary between the disordered fluid phase
and the ordered, polar, phase of the active fluid. We show this boundary in Fig.~\ref{fig:phasediagram}. 
Note that although in this subsection we have neglected all spatial derivatives, it is crucial that we have included the contributions of the higher Fourier modes in the orientational angle, i.e., $\bm{j}_2$. The coupling of this mode to the lower Fourier modes introduces nonlinearity in the system, which is necessary to stabilize the ordered polar phase.

The self-consistent approximation is known not to hold for  phase transitions in which fluctuations are strong. 
Fluctuations can shift the transition line in parameter space and, furthermore, can change the nature of the transition.
We thus expect the prediction
that the order-disorder transition in polar active fluids is continuous not to hold---in physical polar active fluids, the transition may be discontinuous.

\subsection{Hydrodynamic equations: coarse-graining to Toner-Tu theory}
Let us now go beyond the assumption of spatial homogeneity and examine the effects of temporally slow, long-wavelength fluctuations in the density and the current for the case of a polar active fluid.
This yields the hydrodynamic theory which, as for any polar active fluid, reduces to a form of Toner-Tu theory~\cite{Toner1995,Farrell2012}.
We proceed via the method introduced in Ref.~\cite{Bertin2009}.
Starting from Eqs.~(\ref{equationsmotionm1}-\ref{equationsmotion0}), we go to the regime in which the order is weak and therefore $P(\bm{r},\theta)$ depends only weakly on $\theta$. 
Physically, this is the regime in which the active flow is much slower than the microscopic speed of each particle. 
Mathematically, this regime allows us to discard all Fourier modes higher than $\bm{j}_2$. [NB: For the sake of generality, let us also include the parameter $\lambda$ from Eq.~(\ref{lambda_fokker}) which controls the feedback between the translational noise and the angle.] 
Then, we consider the equations
\begin{subequations}
\begin{align}
 \partial_t\rho &=-\frac{1}{2} \nabla\cdot\left(\bm{j}-\frac{1}{\kappa}\nabla \rho\right), \\
\partial_t j_x &= \left(\frac{1}{2\kappa}\nabla^2-m\right) j_x+\left(\frac{\lambda-2}{2}\right)\partial_x \rho, \\
&-\left(\frac{\lambda+2}{4}\right)\nabla \cdot \bm{j}_2+g\left( h_x\rho-\frac{1}{2}\bm{h}\cdot\bm{j}_2\right), \\
 \partial_t j_y& = \left(\frac{1}{2\kappa}\nabla^2-m\right) j_y+\left(\frac{\lambda-2}{2}\right)\partial_y \rho, \\
&-\left(\frac{\lambda+2}{4}\right)\nabla\cdot \bm{j}_{2\perp}+g\left( h_x\rho-\frac{1}{2}\bm{h}\cdot\bm{j}_{2 \perp}\right), \\
\partial_t j_{2,x}&= \left(\frac{1}{2\kappa}\nabla^2-4m\right) j_{2,x}+\left(\frac{\lambda-1}{2}\right)\nabla\ast\bm{j}+ g\bm{h}\ast\bm{j}, \\
\partial_t j_{2,y}&= \left(\frac{1}{2\kappa}\nabla^2-4m\right) j_{2,y}+\left(\frac{\lambda-1}{2}\right)\nabla\ast\bm{j}_{\perp}+g\bm{h}\ast\bm{j_{\perp}}.
\label{equationsmotion}
\end{align}
\end{subequations}
Note two important aspects of the associated approximation: (i)~$|\bm{j}_2|$ is much smaller than $|\bm{j}|$; and (ii)~as we are interested in hydrodynamics, we only consider time- and length-scales much larger than the microscopic ones. Therefore, we consider the regime characterized by
\begin{equation}
\partial_t\bm{j}_{2,x}, \partial_t\bm{j}_{2,y} \ll m\bm{j}_{2,x}, m\bm{j}_{2,y}
\end{equation}
in which both the time-derivatives $\partial_t \bm{j}_2$ and the term $\frac{1}{\kappa}\nabla^2 \bm{j}_2$ are neglected. 
Re-expressing the current $\bm{j}$ via $\bm{j}=\bm{h}/\pi N$ and $\rho_d=2 \pi N \rho$, we can rewrite the above equations as
\begin{equation}
\partial_t \rho_d = -\nabla\cdot\left(\bm{h}-\frac{1}{2\kappa}\nabla\rho_d\right),
\label{continuity}
\end{equation}
and
\begin{align}
\partial_t \bm{h} &=\left[\left(\frac{\rho_d g}{2}-m\right)-\frac{g^2}{8 m} h^2\right] \bm{h}+\left(\frac{\lambda-2}{4}\right)\nabla \rho_d \nonumber \\
& - (\lambda-1)\frac{g}{16m} \left[(\bm{h}\cdot\nabla)\bm{h}+(\bm{h}_{\perp}\cdot\nabla)\bm{h}_{\perp}\right] \nonumber\\ 
&- (\lambda+2)\frac{g}{8m}\left[\bm{h} (\nabla\cdot \bm{h})-\bm{h}_{\perp}(\nabla\cdot\bm{h}_{\perp})\right] \nonumber\\
&+\left[\frac{1}{2 \kappa}+\frac{2-\lambda-\lambda^2}{32 m}\right]\nabla^2 \bm{h}.
\end{align}
Finally, the unusual $\bm{h}_{\perp}$ terms may be rewritten in a more familiar way via the identities
\begin{align}
\bm{h}_{\perp}(\nabla\cdot\bm{h}_{\perp}) &=\frac{1}{2}\nabla h^2-(\bm{h}\cdot\nabla)\bm{h}, \nonumber\\
(\bm{h}_{\perp}\cdot\nabla)\bm{h}_{\perp} &=\frac{1}{2}\nabla h^2-\bm{h}(\nabla\cdot\bm{h}),
\end{align}
where, as above, $h\equiv|\bm{h}|$.
The resulting hydrodynamic equations read:
\begin{align}
\partial_t \bm{h} &+ (\lambda+1)\frac{3g}{16m}(\bm{h}\cdot \nabla)\bm{h}= \left[\left(\frac{\rho_d g}{2}-m\right)-\frac{g^2}{8 m} h^2\right] \bm{h} \nonumber\\
&+\left(\frac{\lambda-2}{4}\right)\nabla \rho_d 
 +(\lambda+5)\frac{g}{16m} \left[\frac{1}{2}\nabla h^2-\bm{h}(\nabla\cdot\bm{h})\right] \nonumber\\ 
&+\left[\frac{1}{2 \kappa}+\frac{2-\lambda-\lambda^2}{32 m}\right]\nabla^2 \bm{h}. 
\label{hydrodynamics}
\end{align} 
Equation~(\ref{continuity}) is the continuity equation that established the conservation of the number of self-propelled particles.  
On the other hand, Eq.~(\ref{hydrodynamics}) describes the hydrodynamics of the polarization order parameter for the active fluid.
Note that the order parameter $\bm{h}$ is not identical to the current of the active fluid, which in addition includes the part of the motion due to translational diffusion.

Let us now discuss the result for the hydrodynamics, Eq.~(\ref{hydrodynamics}). The second term on the left-hand side is an advection term. In general, as consequence of the breaking of Galilean invariance in our system, its prefactor is not unity. This results from the presence of a special reference frame in the system: there is only one frame in which the particles can propagate with equal speed, regardless of their direction of movement. On the right-hand side, the first term is the one responsible for the spontaneous breaking of the rotational symmetry of the fluid in the polar state. The second and third terms are pressure-like terms. The third term includes the effects of a non-linear compressibility. 

\begin{figure}
\centering
\includegraphics[angle=0]{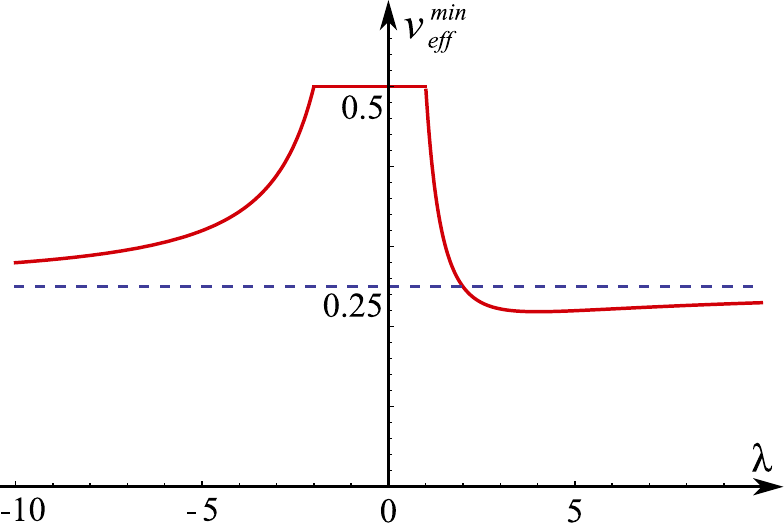}
\caption{The minimum effective viscosity $\nu^{\textit{min}}_{\textit{eff}}$ (in units of $1/\kappa$) possible in the hydrodynamic equation, as a function of the (dimensionless) parameter $\lambda$. The parameter $\lambda$ controls the coupling between orientational and translational noises. Note that the effective viscosity is always positive, and the minimal value has a plateau for $-2\leq \lambda \leq 1$. For $|\lambda| \gg 1$, 
$\nu^{\textit{min}}_{\textit{eff}}$ asymptotes to $0.25$.  For $\lambda > 0$, this asymptote is approached from below, whereas for $\lambda < 0$ the asymptote is approached from above. This plot shows how the coupling between orientational and translational noise of active particles can control the effective viscosity of an active fluid.}
\label{minviscocity}
\end{figure}

Turning now to the fourth term, this is a viscous damping term familiar from, e.g., the Navier-Stokes equations. In the present context, this term includes the effects of both the translational noise and the coupling between the translational and orientational noises. Curiously, the parameter $\lambda$ has a strong effect on the effective viscosity. E.g., if $\lambda=1$ (the model we focus on for the single-particle dynamics), this viscosity reduces to $1/{2\kappa}$. In that case, the viscosity depends only on the translational noise.
At first glance, the expression for the effective viscosity does not appear to be positive-definite. If the viscosity were to change sign, this would suggest that the homogeneous fluid state should become unstable.
However, in the previous section we showed that the particle dynamics is physical only if the inequality $m > \kappa \lambda^2/8$ holds. 
We now re-express the mass in terms of a positive parameter $\epsilon$, via $m=\kappa[\epsilon+(\lambda^2/8)]$, and write the effective viscosity as
\begin{equation}
\label{effviscosity}
\nu_{\textit{eff}} = \frac{1}{2\kappa}\left[1+\frac{2-\lambda-\lambda^2}{16 [\epsilon+(\lambda^2/8)]}\right].
\end{equation}  
The effective viscosity as given by Eq.~(\ref{effviscosity}) is positive for $\lambda$ real and $\epsilon$ nonnegative. To see this, note that when $2-\lambda-\lambda^2 <0$, Eq.~(\ref{effviscosity}) yields the minimal possible value for the effective viscosity for each value of $\lambda$ in the limit $\epsilon\rightarrow 0$. This value is always positive and converges to $1/4\kappa$ as $\lambda \rightarrow \pm \infty$. On the other hand, when $2-\lambda-\lambda^2 >0$, the minimum value that the effective viscosity can achieve is $1/2\kappa$.  We plot the minimum value of the effective viscosity for each value of $\lambda$ in Fig.~\ref{minviscocity}.  Curiously, it is also worth noting that in the case $\lambda=-1$ (a noise coupling with the same strength but with the opposite sign of our original model),
the advection term vanishes.

The hydrodynamic equations~(\ref{hydrodynamics}), along with the continuity equation~(\ref{continuity}), are a version of the Toner-Tu equations, first derived in Ref.~\cite{Toner1995} based on symmetry considerations. By contrast, we obtain these equations based on the microscopic single-particle model that we introduced in the previous section along with inter-particle interactions. The coefficients that we obtained thus explicitly depend on the microscopic parameters of the model, which allows not only for the form but also for the precise numerical evaluation for the hydrodynamic coefficients in the Toner-Tu equations. In the special limit $(\kappa,\lambda)\rightarrow(\infty,0)$, with $\lambda^2 \kappa \rightarrow 0$, the hydrodynamic equations that we derive coincide exactly with the ones derived in Ref.~\cite{Farrell2012}.

\section{Conclusions}
Because the model we consider falls within the Toner-Tu universality class, we can immediately conclude that the polar active particles that we consider
do exhibit long-range order despite the two-dimensional character of the system~\cite{Toner1995}. This is in contrast to equilibrium two-dimensional systems with short-ranged interactions, which cannot break a continuous symmetry, such
as the rotational symmetry of the XY model, as embodied in the Mermin-Wagner theorem~\cite{MerminWagner}.
Therefore, the hydrodynamic theory that we obtain, described by Eqs.~(\ref{hydrodynamics}),
differs from the regular Navier-Stokes equations in two crucial ways: (i)~due to the breaking of Galilean invariance via both
activity and momentum exchange characteristic of dry active matter, the hydrodynamic theory includes terms prohibited in the Navier-Stokes equations; and 
(ii)~as a result of these additional terms, interacting self-propelled particles exhibit long-range polar order.

To summarize, we have introduced a model of active particles based on an analogy with a Schr\"{o}dinger equation that describes the propagation
of an electron subject to spin-orbit coupling. 
We show that this model has a standard description as a stochastic process in terms of either a Fokker-Planck or a Langevin equation, both of which we derive. 
We note that within this stochastic interpretation, the orientational and the translational noise of the active particles we consider
are coupled via a relation reminiscent of the Heisenberg uncertainty principle.
Based on this single-particle physics, we derive a description for a polar active fluid in which the particles preferentially
align their velocities. Within this description, we characterize the transition from a disordered to an ordered (i.e., polar) state
via a hydrodynamic Toner-Tu theory.

Previously, analogies between classical processes and quantum dynamics have found use in areas as diverse as polymer physics, liquid crystal elasticity,
hydrodynamics, and financial markets~\cite{Kleinert}. Such analogies are often drawn via a path-integral formalism but they may instead be formulated by rotating
the time axis into the complex plane. We have shown that this latter approach can be extended to the study 
active fluids composed of self-propelled particles. In order to account for the self-propulsion, we employ concepts familiar from
the study of correlated electron fluids. 
These connections have the potential to help uncover novel phases of active matter via analogies with electronic counterparts.

\section{Acknowledgments}
We thank D.~Bartolo, J.~Toner, V.~Vitelli, and R.~Hipolito for insightful discussions. B.L. would like to thank
CONICYT and Becas Chile for their financial support. A.S. acknowledges funding from the Delta Institute for Theoretical Physics. P.M.G. acknowledges for its hospitality the Aspen Center for Physics, which is supported by National Science foundation grant PHY-1607611. 

\end{document}